\documentclass[10pt,conference,twocolumn]{IEEEtran}

\usepackage{amsmath}
\usepackage{amsfonts}
\usepackage[noend]{algorithmic}
\usepackage{array}
\usepackage{stfloats}
\usepackage{url}

\usepackage{tikz}
\usetikzlibrary{chains,shapes}
\usepackage{pgfplots}

\usepackage{stmaryrd} 
\newcommand{\lsem}{\llbracket}
\newcommand{\rsem}{\rrbracket}

\newcommand{\Def}[1]{\emph{{#1}}}

\newcommand{\BB}{\mathbb{B}}
\newcommand{\LStar}{\texttt{L\textsuperscript{*}}}
\newcommand{\TTT}{\texttt{TTT}}
\newcommand{\bigO}{\mathcal{O}}

\newtheorem{definition}{Definition}
\newtheorem{lemma}{Lemma}

\begin{document}
\title{Learning Product Automata\\{\Large Revisiting a trick by Rivest and Schapire}}

\author{\IEEEauthorblockN{Joshua Moerman}
\IEEEauthorblockA{Institute for Computing and Information Sciences\\
    Radboud University, Nijmegen, the Netherlands\\
    joshua.moerman@cs.ru.nl}
}

\maketitle

\begin{abstract}
In this paper we give an optimization for active learning algorithms,
applicable to learning Moore machines where the output comprises several observables.
These machines can be decomposed themselves by projecting on each observable, resulting in smaller components.
These components can then be learnt with fewer queries.
This is in particular interesting for learning software,
where compositional methods are important for guaranteeing scalability.
\end{abstract}

\section{Introduction}
Active automata learning is becoming a valuable tool in software engineering and verification \cite{VaandragerACM}.
Indeed, applications can be found in a broad range:
finding bugs in network protocols \cite{FiterauTCP},
assisting with refactoring legacy software \cite{SchutsRefactoring},
specification mining \cite{AlurLearningSpecs}, and more.

These learning techniques originate from the field of grammatical inference.
One of the crucial steps for applying these techniques on software was to move from deterministic finite automata to deterministic Moore or Mealy machines, capturing reactive systems with any kind of output.
With little adaptations, the algorithms work well, as shown by the many applications.
This is remarkable, since little software specific knowledge is used (besides the input alphabet of actions).

Realizing that software is often composed of smaller pieces, it makes sense to incorporate such information in learning algorithms.
In the present paper we aim to do exactly that for the simplest case of composition: we learn product automata.

To the best of the author's knowledge, this has not been done before explicitly.
Furthermore, libraries such as LearnLib \cite{LearnLib} and libalf \cite{libalf} do not include such functionality.
\emph{Implicitly}, however, it has been done before.
Rivest and Schapire use two tricks to reduce the size of some automata in their paper ``Diversity-based inference of finite automata'' \cite{RivestSchapireDiversity}.
The first trick is to look at the \emph{reversed automaton} (in their terminology, the \emph{diversity-based automaton}).
The second trick (which is not explicitly mentioned, unfortunately) is to have a different automaton for each observable (i.e. output).
In one of their examples the two tricks combined give a reduction from $\pm 10^{19}$ states to just $54$ states.

We isolate this trick, so we can apply it more generally.
Furthermore, we argue that this is particularly interesting in the context of model learning of software, as composition is a common tool in software engineering.

\section{Preliminaries}

We use the formalism of Moore machines to describe our algorithms.
Nonetheless, the results can also be phrased in terms of Mealy machines.

\begin{definition}
	A \Def{Moore machine} is a tuple $M = (Q, I, O, \delta, o, q_0)$
	where $Q, I$ and $O$ are finite sets of states, inputs and outputs respectively,
	$\delta: Q \times I \to Q$ is the transition function,
	$o: Q \to O$ is the output function,
	and $q_0$ is the initial state.
	The size $|M|$ is the cardinality of $Q$.
\end{definition}

We extend the definition of the transition function to words as $\delta: Q \times I^\ast \to Q$.
The \Def{behaviour} of a state $q$ is the map $\lsem q \rsem: I^\ast \to O$ defined by $\lsem q \rsem(w) = o(\delta(q, w))$.
We extend this to the machine $\lsem M \rsem = \lsem q_0 \rsem$.
Two states $q,q'$ are \Def{equivalent} if $\lsem q \rsem = \lsem q' \rsem$.
Two machines are \Def{equivalent} if their initial states are.
A machine is \Def{minimal} if all states have different behaviour and all states are reachable.

\begin{definition}
	Given two Moore machines with equal input sets $M_1 = (Q_1, I, O_1, \delta_1, o_1, {q_0}_1)$ and $M_2 = (Q_2, I, O_2, \delta_2, o_2, {q_0}_2)$,
	we define their \Def{product} $M_1 \times M_2$ by:
	$$ M_1 \times M_2 = (Q_1 \times Q_2, I, O_1 \times O_2, \delta, o, ({q_0}_1, {q_0}_2)),$$
	where $ \delta((q_1, q_2), a) = (\delta_1(q_1, a), \delta_2(q_2, a)) $ and
	$ o((q_1, q_2)) = (o_1(q_1), o_2(q_2)) $.
\end{definition}

The product is formed by running both machines in parallel and letting $I$ act on both machine simultaneously.
The output of both machines is observed.
Note that the product Moore machine might have unreachable states, even if the components are reachable.
The product of more than two machines is defined by induction.

Let $M$ be a machine with outputs in $O_1 \times O_2$.
By post-composing the output function with projection functions we get two machines, called \Def{components}, $M_1$ and $M_2$ with outputs in $O_1$ and $O_2$ respectively.
Then $M$ is equivalent to $M_1 \times M_2$.
If $M$ and its components $M_i$ are taken to be minimal, then we have $|M| \leq |M_1| \cdot |M_2|$ and $|M_i| \leq |M|$.
In the best case we have $|M_i| = \sqrt{|M|}$ and so the behaviour of $M$ can be described using only $2\sqrt{|M|}$ states, which is less than $|M|$ (if $|M| > 4$).
With iterated products the reduction can be even more as shown in the following example.
This reduction in state-space can be exploited by learning algorithms, as will be shown in later sections.

We introduce basic notation: $\pi_i : A_1 \times A_2 \to A_i$ are the usual projection functions.
On a function $f: X \to A_1 \times A_2$ we use the shorthand $\pi_i f$ to denote $\pi_i \circ f$.
As usual, $uv$ denotes concatenation of string $u$ and $v$, and this is lifted to sets of strings $UV = \{uv \mid u \in U, v \in V\}$.
We define the set $[n] = \{1, \dots, n\}$ and the set of Boolean values $\BB = \{0, 1\}$.

\subsection{Example}
\begin{figure}
	\centering
	\begin{tikzpicture}
	\tikzstyle{bitcell}=[draw,minimum size=0.6cm]
	\tikzstyle{machinehead}=[arrow box,draw,minimum size=.37cm,arrow box arrows={north:.25cm}]
	
	\begin{scope}[start chain=1 going right,node distance=-0.15mm]
	\node [on chain=1,bitcell] {0};
	\node [on chain=1,bitcell] {0};
	\node [on chain=1,bitcell] {1};
	\node [on chain=1,bitcell] (input) {0};
	\node [on chain=1,bitcell] {0};
	\node [on chain=1,bitcell] {1};
	\node [on chain=1,bitcell] {0};
	\node [on chain=1,bitcell] {1};
	\end{scope}
	
	\node [machinehead,yshift=-.3cm] at (input.south) (head) {};
	
	\end{tikzpicture}
	\caption{A state of the $8$-bit machine}
	\label{fig:bitmachine}
\end{figure}
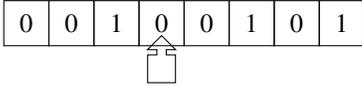

We take the \Def{$n$-bit register machine} example from \cite{RivestSchapireDiversity}.
The state space of the $n$-bit register machine $M_n$ is given by $n$ bits and a position of the reading/writing head, see Figure~\ref{fig:bitmachine}.
The inputs are commands to control the position of the head and to flip the current bit. The output is the current bit vector.
Formally it is defined as $M_n = (\BB^n \times [n], \{L, R, F\}, \BB^n, \delta, o, i)$, where the initial state is $i = ((0, \dots, 0), 1)$, the output is $o(((b_1, \dots, b_n), k)) = (b_1, \dots, b_n)$ and the transition function is given by
\begin{eqnarray*}
	\delta(((b_1, \dots, b_n), k), L) &=& \begin{cases}
	((b_1, \dots, b_n), k-1) \quad &\text{if } k > 1 \\
	((b_1, \dots, b_n), n) \quad &\text{if } k = 1
	\end{cases} \\
	\delta(((b_1, \dots, b_n), k), R) &=& \begin{cases}
	((b_1, \dots, b_n), k+1) \quad &\text{if } k < n \\
	((b_1, \dots, b_n), 1) \quad &\text{if } k = n
	\end{cases} \\
	\delta(((b_1, \dots, b_n), k), F) &=& ((b_1, \dots, \neg b_k, \dots, b_n), k),
\end{eqnarray*}
that is, $L$ moves the head to the left and $R$ to the right (and wraps around on the ends), while $F$ flips the current bit.

The machine $M_n$ is minimal and has $n \cdot 2^n$ states.
So although this machine has very simple behaviour, learning it will require a lot of queries because of its size.
Luckily, the machine can be decomposed into smaller components.
For each bit $l$ we define a component $M_n^l = (\BB \times [n], \{L,R,F\}, \BB, \delta^l, o^l, (0,1))$ where $o^l((b,k)) = b$ and
\begin{eqnarray*}
	\delta^l((b, k), L) &=& \begin{cases}
		(b, k-1) \quad &\text{if } k > 1 \\
		(b, n) \quad &\text{if } k = 1
	\end{cases} \\
	\delta^l((b, k), R) &=& \begin{cases}
		(b, k+1) \quad &\text{if } k < n \\
		(b, 1) \quad &\text{if } k = n
	\end{cases} \\
	\delta^l((b, k), F) &=& \begin{cases}
		(\neg b, k) \quad &\text{if } l = k \\
		(b, k) \quad &\text{if } l \neq k
	\end{cases}
\end{eqnarray*}

The product $M_n^1 \times \dots \times M_n^n$ is equivalent to $M_n$.
Each of the components $M_n^l$ is minimal and has only $2n$ states.
So by this decomposition, we only need $2 \cdot n^2$ states to describe the whole behaviour of $M_n$.
Note, however, that the product $M_n^1 \times \dots \times M_n^n$ is not minimal: many states are unreachable.

\section{Learning}

We describe two approaches for active learning of product machines.
One is a direct extension of the well-known \LStar{} algorithm.
The other reduces the problem to any active learning algorithm, so that one can use more optimised algorithms.

We fix an unknown target machine $M$ with a known input alphabet $I$ and output alphabet $O = O_1 \times O_2$.
The goal of the learning algorithm is to infer a machine equivalent to $M$, given access to a \Def{minimally adequate teacher} \cite{Angluin}.
The teacher will answer two types of queries:
\begin{itemize}
	\item \Def{Membership queries} $MQ(w)$ for words $w \in I^\ast$, the teacher will answer with $\lsem M \rsem (w) \in O$.
	\item \Def{Equivalence queries} $EQ(M')$ for a machine $M'$ on the same alphabets, the teacher will answer with \texttt{YES} if $M$ and $M'$ are equivalent and she will answer with a word $w$ such that $\lsem M \rsem (w) \neq \lsem M' \rsem (w)$ otherwise.
\end{itemize}

\subsection{Learning product automata with an \LStar{} extension}

We can use the general framework for automata learning as set up in \cite{vHeerdtCALF,vHeerdtThesis}.
The general account does not directly give concrete algorithms, but it does give generalised definitions for \Def{closedness} and \Def{consistency}.
The main data structure for the algorithm is an observation table.

\begin{definition}
	An \Def{observation table} is a triple $(S, E, T)$ where $S, E \subseteq I^\ast$ are finite sets of words and $T: S \cup SI \to O^E$ is defined by $T(s)(e) = \lsem M \rsem (se)$.
\end{definition}

During the \LStar{} algorithm the sets $S,E$ grow and $T$ encodes the knowledge of $\lsem M \rsem$ so far.

\begin{definition}
	An observation table $(S,E,T)$ is \Def{product-closed} if for all $t \in SI$ there exist $s_1, s_2 \in S$ such that $\pi_i T(t) = \pi_i T(s_i)$ for $i=1,2$.
	It is \Def{product-consistent} if for $i=1,2$ and for all $s, s' \in S$ we have $\pi_i T(s) = \pi_i T(s')$ implies $\pi_i T(sa) = \pi_i T(s'a)$ for all $a \in I$.
\end{definition}

These definitions are related to the classical definitions of closedness and consistency.
In fact the latter two points of the following lemma restate the above definitions.
For the first two points the converse does not necessarily hold.

\begin{lemma}
	\label{lemma:implications}
	Let $OT = (S,E,T)$ be an observation table and let $\pi_i OT = (S,E,\pi_i T)$ be a component.
	We have the following implications:
	\begin{enumerate}
		\item $OT$ is closed $\implies$ $OT$ is product-closed,
		\item $OT$ is consistent $\impliedby$ $OT$ is product-consistent,
		\item $OT$ is product-closed $\iff$ $\pi_i OT$ is closed $\forall i$,
		\item $OT$ is product-consistent $\iff$ $\pi_i OT$ is consistent $\forall i$.
	\end{enumerate}
\end{lemma}

\begin{lemma}
	\label{lemma:construct}
	Given a product-closed and -consistent table we can define a product Moore machine consistent with the table, where each component is minimal.
\end{lemma}

\begin{figure}
	\begin{algorithmic}[1]
		\STATE Initialise $S$ and $E$ to $\{ \epsilon \}$
		\STATE Initialise $T$ with MQs
		\REPEAT
		\WHILE{$(S,E,T)$ is not product-closed or -consistent}
		\IF{$(S,E,T)$ not product-closed}
		\STATE find $t \in SI$ such that there is no $s \in S$ with $\pi_i T(t) = \pi_i T(s)$ for some $i$
		\STATE add $t$ to $S$ and fill the new row using MQs
		\ENDIF
		\IF{$(S,E,T)$ not product-consistent}
		\STATE find $s, s' \in S$, $a \in I$ and $e \in E$ such that $\pi_i T(s) = \pi_i T(s')$ but $\pi_i T(sa)(e) \neq \pi_i T(s'a)(e)$ for some $i$
		\STATE add $ae$ to $E$ and fill the new column using MQs
		\ENDIF
		\ENDWHILE
		\STATE Construct $H$ (by Lemma~\ref{lemma:construct})
		\IF{$EQ(H)$ gives a counterexample $w$}
		\STATE add $w$ and all its prefixes to $S$
		\STATE fill the new rows with MQs
		\ENDIF
		\UNTIL{$EQ(H) =$ \texttt{YES}}
		\RETURN $H$
	\end{algorithmic}
	\caption{The product-\LStar{} algorithm}
	\label{alg:product-lstar}
\end{figure}

We list the product-\LStar{} algorithm in Figure~\ref{alg:product-lstar}.
Its termination follows from the fact that \LStar{} terminates on both components.

By Lemma~\ref{lemma:implications} (1) we note that the algorithm does not need more rows than we would need by running \LStar{} on $M$.
By point (4) of the same lemma, we find that it does not need more columns than \LStar{} would need on each component combined.
This means that in the worst case, the table is twice as big as the original \LStar{} would do.
However, in good cases (such as the running example), the table is much smaller, as the number of rows is less for each component and the columns needed for each component may be similar.

\subsection{Learning product automata via a reduction}

The previous algorithm constructs two machines from a single table.
This suggests that we can also run two learning algorithms to construct two machines.
We lose the fact that the data structure is shared between the learners, but we gain that we can use more efficient algorithms than \LStar{} without any effort.

The crucial observation is that a counterexample is necessarily a counterexample for at least one of the two learners.
In this case we simply forward the counterexample to that learner.
(If at a certain stage only one learner makes an error, we keep the other learner suspended, as we may obtain a counterexample for that one later on.)
This observation means that at least one of the learners makes progress and will eventually terminate.
Hence, the whole algorithm will terminate.

In the worst case, twice as many queries will be posed, compared to learning the whole machine at once.
(This is because learning the full machine also learns its components.)
In good cases, such as the running example, it requires much less queries.
Typical learning algorithms require $\bigO(n^2)$ membership queries ($n$ being the number of states of the minimal machine).
For the example $M_n$ this gives $\bigO((n \cdot 2^n)^2) = \bigO(n^2 \cdot 2^{2n})$ queries.
When learning the components $M_n^l$ with the above algorithm, that gives just $\bigO((2n)^2 + \dots + (2n)^2) = \bigO(n^3)$ queries.

\begin{figure}
	\begin{algorithmic}[1]
		\STATE Initialise two learners $L_1$ and $L_2$
		\REPEAT
		  \WHILE{$L_i$ queries $MQ(w)$}
		    \STATE forward $MQ(w)$ to the teacher and get output $o$
		    \STATE return $\pi_i o$ to $L_i$ 
		  \ENDWHILE
		  \COMMENT{at this point both learners constructed a hypothesis}
		  \STATE Let $H_i$ be the hypothesis of $L_i$
		  \STATE Construct $H = H_1 \times H_2$
		  \IF{$EQ(H)$ returns a counterexample $w$}
		    \IF{$\lsem H_1 \rsem (w) \neq \pi_1 \lsem M \rsem (w)$}
		      \STATE return $w$ to $L_1$
		    \ENDIF
		    \IF{$\lsem H_2 \rsem (w) \neq \pi_2 \lsem M \rsem (w)$}
		      \STATE return $w$ to $L_2$
		    \ENDIF
		  \ENDIF
		\UNTIL{$EQ(H) =$ \texttt{YES}}
		\STATE return \texttt{YES} to both learners
		\RETURN $H$
	\end{algorithmic}
	\caption{Learning product machines with other learners}
	\label{alg:general-product-learner}
\end{figure}

\section{Experiments}

The algorithm via reduction is implemented in LearnLib.\footnote{
	The implementation and models can be found on-line at\\
	\url{https://gitlab.science.ru.nl/moerman/learning-product-automata}}
As we expect the reduction algorithm to be the most efficient (as it can use an efficient learner internally), we leave an implementation of the direct extension of \LStar{} as future work.
The implementation handles products of any size (as opposed to only products of two machines).

In this section we compare the product learner with a regular learning algorithm (we use the \TTT{} algorithm \cite{IsbernerTTT} for the comparison).
We measure the number of equivalence queries and membership queries.
In addition, the equivalence queries are implemented by random sampling so as to imitate the intended application of learning black-box software.
Efficiency can then be measured by the total number of input actions sent to the machine (including resets).
The results can be found in Table~\ref{tab:results}.
We have two sets of models.

\paragraph{$n$-bit register machine}
The machines $M_n$ are as described before.
We note that the product learner is much more efficient, as expected.

\paragraph{Circuits}
In addition to the (somewhat artificial) examples $M_n$, we use circuits which appeared in the logic synthesis workshops (LGSynth89/91/93), part of the ACM/SIGDA benchmarks.\footnote{
	The original files describing these circuits can be found at\\
	\url{https://people.engr.ncsu.edu/brglez/CBL/benchmarks/}}
These models have been used as benchmarks before for FSM-based testing methods \cite{HieronsTurkerCircuits} and describe the behaviour of real-world circuits.
The circuits have bit vectors as outputs, and can hence be naturally be decomposed by taking each bit individually.
For the circuit \emph{mark1}, we did not split the $16$-bit output to individual bits.
Instead, we grouped the bits in pairs, resulting in $8$ components.

For some but not all circuits the number of membership queries is reduced compared to a regular learner.
Unfortunately, the results are not as impressive as for the $n$-bit register machine.
We do note, however, that in all cases the number of actions needed in total is reduced.

\begin{figure}
	\begin{tikzpicture}
	\begin{axis}[
	height=5cm,
	width=\columnwidth,
	xmin=0.1,
	xmax=22.9,
	xlabel={Hypothesis},
	ylabel={number of states},
	ymode=log
	]
	
	\addplot[black, mark=+] plot coordinates {
		(1,   1)
		(2,   8)
		(3,   7)
		(4,   25257)
		(5,   1033)
		(6,   2312)
		(7,   855)
		(8,   896)
		(9,   615)
		(10,  865)
		(11,  55230)
		(12,  7023)
		(13,  16235)
		(14,  58128)
		(15,  1629)
		(16,  2782)
		(17,  941)
		(18,  220)
		(19,  200)
		(20,  204)
		(21,  204)
		(22,  202)
	};
	\end{axis}
	\end{tikzpicture}
	\caption{The number of states for each hypothesis while learning \emph{mark1}}
	\label{fig:stategrowth}
\end{figure}
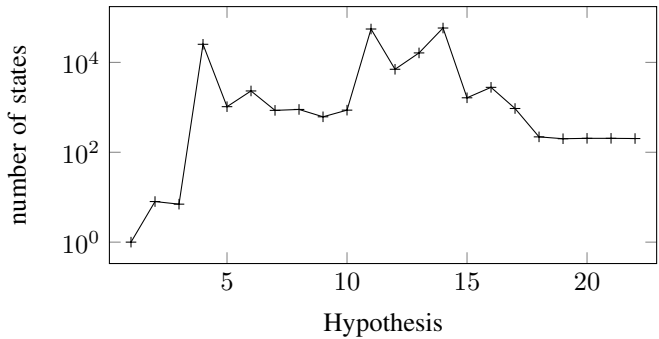

In Figure~\ref{fig:stategrowth}, we look at the size of each hypothesis generated during the learning process.
We note that, although each component grows monotonically, the number of reachable states in the product does not grow monotonically.
In this particular instance where we learn \emph{mark1} there was a hypothesis of $58\,128$ states, much bigger than the target machine of $202$ states.
In the theoretical framework, this is not an issue, as the teacher will allow it and answer the query.
Even in the PAC model, this poses no problem as we can efficiently determine membership.
However, in some applications the equivalence queries are implemented with a model checker or some sophisticated test generation tool \cite{FiterauTCP}.
In these cases, the increased hypotheses may be undesirable.

\begin{table*}
	\centering
	\begin{tabular}{l|r r|r r r|r r r|}
		& & & \multicolumn{3}{c}{Product learner} & \multicolumn{3}{c}{\TTT{} learner} \\
		Machine & States & Components & EQs & MQs & Actions & EQs & MQs & Actions \\ 
		\hline
		$M_2$ & $8$    & $2$ & $3$  & $100$    & $621$     & $5$   & $115$     & $869$      \\
		$M_3$ & $24$   & $3$ & $3$  & $252$    & $1\,855$  & $5$   & $347$     & $2946$     \\
		$M_4$ & $64$   & $4$ & $8$  & $456$    & $3\,025$  & $6$   & $1\,058$  & $13\,824$  \\
		$M_5$ & $160$  & $5$ & $6$  & $869$    & $7\,665$  & $17$  & $2\,723$  & $34\,657$  \\
		$M_6$ & $384$  & $6$ & $11$ & $1\,383$ & $12\,870$ & $25$  & $6\,250$  & $90\,370$  \\
		$M_7$ & $896$  & $7$ & $11$ & $2\,087$ & $24\,156$ & $52$  & $14\,627$ & $226\,114$ \\
		$M_8$ & $2048$ & $8$ & $13$ & $3\,289$ & $41\,732$ & $160$ & $34\,024$ & $651\,678$ \\
		\hline
		bbara  & $7$   & $2$ & $3$  & $167$     & $1\,049$   & $3$  & $216$     & $1\,535$   \\
		mark1 & $202$ & $8^*$ & $22$ & $13\,027$ & $117\,735$ & $67$ & $15\,192$ & $252\,874$ \\
		keyb   & $41$  & $2$ & $25$ & $12\,464$ & $153\,809$ & $24$ & $6024$    & $265\,805$ \\
		ex3    & $28$  & $2$ & $24$ & $1\,133$  & $9\,042$   & $18$ & $878$     & $91\,494$
	\end{tabular}
	\caption{Comparison of the product learner with an ordinary learner}
	\label{tab:results}
\end{table*}

\section{Final remarks}

We have shown two learning algorithms which exploit a decomposable output.
If the output can be split, then also the machine itself can be decomposed in components.
As the few experiments show, this can be a very effective optimization for learning software.
It should be stressed that the improvement of the optimization depends heavily on the independence of the components.
For example, the $n$-bit register machine has nearly independent components and the reduction in the number of queries is big.
The more realistic circuits did not show such improvements.
A potential problem is the growth of the intermediate hypotheses.
In the remainder of this section we discuss related ideas left for future work.

\subsection{Generalization to subsets of products}

In some cases we might know even more about our output alphabet.
The output set $O$ may be a proper subset of $O_1 \times O_2$, indicating that some outputs can only occur ``synchronised''.
For example, we might have $O=\{ (0,0) \} \cup \{ (a,b) \mid a,b \in [3] \}$, i.e. the output $0$ for either component can only occur if the other component is also $0$.

In such cases we can use the above algorithm still, but we may insist that the teacher only accepts machines with output in $O$ for the equivalence queries (as opposed to outputs in $\{0,1,2,3\}^2$).
When constructing $H = H_1 \times H_2$ in line 7 of Figure~\ref{alg:general-product-learner}, we can do a reachability analysis on $H$ to check for non-allowed outputs.
If such traces exist, we know it is a counterexample for at least one of the two learners.
With such traces we can fix the defect ourselves, without having to rely on the teacher.

\subsection{The other trick by Rivest and Schapire}

The main trick of \cite{RivestSchapireDiversity} was to exploit the structure of the so-called ``diversity-based'' automaton.
This automaton may also be called the reversed Moore machine.
It provides a duality between reachability and equivalence which is the core idea of Brzozowski's minimization algorithm \cite{RotReverseMoore,BonchiReverseMoore}.

Let $M^R$ denote the reverse of $M$, then we have $\lsem M^R \rsem (w) = \lsem M \rsem (w^R)$.
This allows us to give an \LStar{} algorithm which learns $M^R$ by posing membership queries with the words reversed.
We computed $M^R$ for the circuit models and all but one of them was much larger than the original.
This suggests that it might not be useful as a trick in learning software, however, a more thorough investigation is desired.

\subsection{Other types of composition}

In this paper, we only looked at the simplest type of composition: products of outputs.
On the other side we can look at combining inputs.
By taking the disjoint union of inputs sets $I_1$ and $I_2$ we can run two machines in parallel, the first is acted upon by $I_1$ and the second by $I_2$.
In this simple construction the machines are fully parallel.
That is, the inputs from $I_1$ commute with the inputs from $I_2$.
More generally, it is interesting to investigate what happens if there are interactions.

A more complex type of product is the cascaded product (also called the wreath product) where the transition structure of the second component may depend on the output (or even state) of the first component.
This captures hierarchical composition \cite{NehanivHierarchical}.
The connection of these more general compositions and learning is left as future work.


\end{document}